\documentclass[12pt,preprint]{aastex}
\shorttitle{BINARY WHITE DWARFS}
\shortauthors{KILIC ET AL.}

\begin{document}

\title{THE DISCOVERY OF BINARY WHITE DWARFS THAT WILL MERGE WITHIN 500 MYR\footnote{Based
on observations obtained at the MMT Observatory, a joint facility of
the Smithsonian Institution and the University of Arizona.}}

\author{Mukremin Kilic\altaffilmark{2,7},
Warren R. Brown\altaffilmark{2},
Carlos Allende Prieto\altaffilmark{3,4}, 
S. J. Kenyon\altaffilmark{2}, and
J. A. Panei\altaffilmark{5,6}
}

\altaffiltext{2}{Smithsonian Astrophysical Observatory, 60 Garden St., Cambridge, MA 02138, USA}
\altaffiltext{3}{Mullard Space Science Laboratory, University College London, Holmbury St. Mary, Surrey RH5 6NT, UK}
\altaffiltext{4}{Current Address: Instituto de Astrof\'{\i}sica de Canarias, 38205 La Laguna, Tenerife, Spain}
\altaffiltext{5}{Facultad de Ciencias Astron\'omicas y Geof\'isicas, UNLP, Paseo del Bosque S/N, La Plata B1900FWA, Argentina}
\altaffiltext{6}{Instituto de Astrof\'isica La Plata, IALP, CONICET-UNLP}
\altaffiltext{7}{{\it Spitzer Fellow}; mkilic@cfa.harvard.edu}

\begin{abstract}

We present radial velocity observations of four extremely low-mass ($0.2~M_\odot$) white dwarfs.
All four stars show peak-to-peak radial velocity variations of 540 -- 710 km s$^{-1}$ with 1.0 -- 5.9 hr
periods. The optical photometry rules out main-sequence companions. In addition, no milli-second pulsar companions
are detected in radio observations. Thus the invisible companions are most likely white dwarfs.
Two of the systems are the shortest period binary white dwarfs yet discovered.
Due to the loss of angular momentum through gravitational radiation, three of the systems will merge within
500 Myr. The remaining system will merge within a Hubble time. The mass functions for three of the systems
imply companions more massive than $0.46~M_\odot$; thus those are carbon/oxygen core white dwarfs.
The unknown inclination angles prohibit a definitive conclusion about the future
of these systems. However, the chance of a supernova Ia event is only 1\% to 5\%.
These systems are likely to form single R Coronae Borealis stars, providing evidence for a white dwarf + white dwarf
merger mechanism for these unusual objects. One of the systems,
SDSS J105353.89+520031.0 has a 70\% chance of having a low-mass white dwarf companion. This system will probably
form a single helium-enriched subdwarf O star.
All four white dwarf systems have unusal mass ratios of $\leq0.2-0.8$ that may also lead to
the formation of AM CVn systems. 

\end{abstract}


\section{INTRODUCTION}

Mergers of binary white dwarfs (WDs) have been proposed to explain supernovae (SNe) Ia events, extreme helium stars including
R Coronae Borealis (RCrB) stars, and single subdwarf B and O stars \citep{iben84,webbink84,saio00,saio02,heber09}.
However, radial velocity surveys of WDs have not revealed a large binary population
that will merge within a Hubble time
\citep{marsh95,maxted00,napiwotzki01,napiwotzki02,karl03,nelemans05}.
In addition to a few pre-WD + WD merger systems \citep[e.g.][]{geier07, tovmassian10},
\citet{napiwotzki04a} identify only eight merger candidates
from the SN Ia Progenitor Survey (SPY) and the literature.

Radial velocity observations of extremely low-mass (ELM, $M\sim0.2~M_\odot$)
WDs provide a new opportunity to find short period binaries.
The Universe is not old enough to produce ELM WDs through
single star evolution. These WDs must therefore undergo significant mass loss during their formation in
binary systems. The majority of ELM WDs have been identified as companions to milli-second pulsars.
However, not all ELM WDs have such companions \citep{vanleeuwen07,agueros09a}.
Radial velocity, radio, and x-ray observations of the lowest gravity WD found in the
Sloan Digital Sky Survey (SDSS), J0917+4638, show that the companion is almost certainly another
WD \citep{kilic07a,kilic07b,agueros09b}.

Recently, \citet{eisenstein06} identified a dozen ELM WDs in the SDSS Data Release 4 area.
Here, we present radial velocity observations of four WDs from that sample;
SDSS J082212.57+275307.4, SDSS J084910.13+044528.7, SDSS J105353.89+520031.0, and SDSS J143633.29+501026.8.
Our observations are discussed in Section 2; an analysis of the spectroscopic data and the nature of
the companions are discussed in Section 3. The future of these binary systems and the merger products are discussed in Section 4.

\section{OBSERVATIONS}

We used the 6.5m MMT equipped with the Blue Channel Spectrograph to obtain moderate
resolution spectroscopy of four ELM WDs on UT 2009 March 27-29 and April 1-3. In addition, we observed J0822+2753
on UT 2008 September 23-24.
We used a 1$\arcsec$ slit and the 832 line mm$^{-1}$ grating in second order to obtain spectra with a wavelength
coverage of 3600 $-$ 4500 \AA\ and a resolving power of $R=$ 4300.
We obtained all spectra at the parallactic angle and acquired He-Ne-Ar comparison lamp exposures after every science exposure.
We checked the stability of the spectrograph by measuring the centroid of the Hg emission line at 4358.34\AA.
The line is stable to within 10 km s$^{-1}$, with an average offset from the rest wavelength of $-0.4 \pm 4.9$ km s$^{-1}$.
We flux-calibrated the spectra using blue spectrophotometric standards \citep{massey88}.

\citet{brown06} observed J1053+5200 in 2006 February as part of their hypervelocity B-star survey.
We include this additional MMT spectrum in our analysis to extend the time baseline.
We use the cross-correlation package RVSAO \citep{kurtz98} to measure heliocentric radial velocities.
We obtain preliminary velocities by cross-correlating the observations with bright WD templates of known velocity.
However, greater velocity precision comes from cross-correlating the objects with themselves.
Thus we shift the individual spectra to rest-frame and sum them together into a high signal-to-noise ratio template spectrum for each
object.
Our final velocities come from cross-correlating the individual observations with these templates,
and are presented in Table 1.

%TABLE1
\begin{deluxetable}{lcr}
\tabletypesize{\footnotesize}
\tablecolumns{3}
\tablewidth{0pt}
\tablecaption{Radial Velocity Measurements}
\tablehead{
\colhead{Object}&
\colhead{HJD}&
\colhead{Heliocentric Radial Velocity}\\
 & +2450000 & (km s$^{-1}$)
}
\startdata
J0822+2753 & 4732.98282 & $-$254.7 $\pm$ 13.6 \\
\nodata    & 4732.99047 & $-$303.4 $\pm$ 11.9 \\
\nodata    & 4733.00223 & $-$312.4 $\pm$ 13.5 \\
\nodata    & 4733.97631 & $-$295.7 $\pm$ 13.4 \\
\nodata    & 4733.99015 & $-$238.0 $\pm$ 14.1 \\
\nodata    & 4917.61018 & +216.2 $\pm$ 10.8 \\
\nodata    & 4922.62095 & $-$291.8 $\pm$  4.5 \\
\nodata    & 4922.68488 &  +99.5 $\pm$  4.8 \\
\nodata    & 4922.75956 &  +97.1 $\pm$  7.6 \\
\nodata    & 4922.79519 & $-$121.5 $\pm$  7.8 \\
\nodata    & 4923.62620 & $-$148.0 $\pm$  5.3 \\
\nodata    & 4923.66669 & +128.6 $\pm$  6.4 \\
\nodata    & 4923.70522 & +223.9 $\pm$  7.3 \\
\nodata    & 4923.76270 &  $-$60.8 $\pm$  5.9 \\
\nodata    & 4924.62160 &  $-$37.1 $\pm$  8.3 \\
\nodata    & 4924.74512 &  $-$99.8 $\pm$ 11.2 \\
\\ \hline \\
J0849+0445 & 4918.64083 &  +415.3   $\pm$ 11.0 \\
\nodata    & 4922.64106 &  +247.1   $\pm$ 14.1 \\
\nodata    & 4922.70390 &  $-$195.9 $\pm$ 10.4 \\
\nodata    & 4922.77816 &  $-$283.1 $\pm$ 11.2 \\
\nodata    & 4923.64966 &  $-$144.7 $\pm$ 12.9 \\
\nodata    & 4923.73361 &  $-$12.6  $\pm$ 10.9 \\
\nodata    & 4923.78104 &  $-$136.6 $\pm$ 12.7 \\
\nodata    & 4924.64406 &  $-$34.2  $\pm$ 11.0 \\
\nodata    & 4924.66382 &  $-$292.4 $\pm$ 13.5 \\
\nodata    & 4924.68169 &  +189.2   $\pm$ 18.8 \\
\nodata    & 4924.70222 &  +391.2   $\pm$ 10.3 \\
\nodata    & 4924.72610 &  $-$142.9 $\pm$ 14.9 \\
\\ \hline \\
J1053+5200 & 3790.79575 &  $-$45.2  $\pm$  8.5 \\
\nodata    & 4919.62596 &   +287.6  $\pm$  8.9 \\
\nodata    & 4922.66788 &  $-$261.0 $\pm$  9.3 \\
\nodata    & 4922.74298 &  +98.6    $\pm$  10.4 \\
\nodata    & 4922.82846 &  +10.8    $\pm$  7.4 \\
\nodata    & 4922.91686 &  $-$65.2  $\pm$  7.8 \\
\nodata    & 4923.68895 &  $-$220.1 $\pm$  7.8 \\
\nodata    & 4923.80370 &  +177.3   $\pm$  7.7 \\
\nodata    & 4923.87868 &  +220.5   $\pm$  6.4 \\
\nodata    & 4924.76454 &  $-$45.3  $\pm$  14.0 \\
\nodata    & 4924.78045 &  +260.2   $\pm$  13.7 \\
\nodata    & 4924.79571 &  $-$249.6 $\pm$  14.7 \\
\nodata    & 4924.81573 &  +235.9   $\pm$  13.5 \\
\nodata    & 4924.83218 &  $-$60.0  $\pm$  13.5 \\
\\ \hline \\
J1436+5010 & 4922.85253 &  +301.4   $\pm$  7.4 \\
\nodata    & 4922.93341 &  $-$50.8  $\pm$  11.5 \\
\nodata    & 4922.98659 &  +224.2   $\pm$  5.6 \\
\nodata    & 4923.82133 &  +230.8   $\pm$  6.8 \\
\nodata    & 4923.85329 &  +89.8    $\pm$  6.7 \\
\nodata    & 4923.89751 &  +41.9    $\pm$  5.8 \\
\nodata    & 4923.93904 &  $-$188.1 $\pm$  7.0 \\
\nodata    & 4923.98171 &  $-$296.2 $\pm$  6.6 \\
\nodata    & 4923.99118 &  +109.6   $\pm$  6.1 \\
\nodata    & 4924.94756 &  $-$140.7 $\pm$  9.3 \\
\nodata    & 4924.95335 &  +112.2   $\pm$  11.8 \\
\nodata    & 4924.96121 &  +327.2   $\pm$  10.6 \\
\nodata    & 4924.96700 &  +243.6   $\pm$  15.9 \\
\nodata    & 4924.97398 &  $-$72.7  $\pm$  11.0 \\
\nodata    & 4924.97975 &  $-$324.1 $\pm$  11.0 \\
\nodata    & 4924.98684 &  $-$372.5 $\pm$  14.0 \\
\enddata
\end{deluxetable}

We also use WD model spectra with atmospheric parameters customized for each object (see \S 3) to
measure radial velocities. The results are consistent within 10 km s$^{-1}$.
Thus, the systematic errors in our measurements are $\leq10$ km s$^{-1}$;
the mean velocity difference between the analyses is $1.5 \pm 3.7$ km s$^{-1}$.
This small uncertainty gives us confidence
that the velocities in Table 1 are reliable.

\section{RESULTS}

All four targets display radial velocity variations of $\geq 540$ km s$^{-1}$ between different observations.
We weight each velocity by its associated error and solve for the best-fit
orbit using the code of \citet{kenyon86}. The heliocentric radial velocities are best fit
with circular orbits and with velocity semi-amplitudes $K= 265 - 367$ km s$^{-1}$.
The best-fit orbital periods range from 0.0426 to 0.2440 days (1.0 to 5.9 hr).
Figures 1-4 show the observed radial velocities and the best fit orbits for our targets.
We have velocity data from $3-6$ nights spread over a time baseline of $3-1134$ nights.
The short orbital periods and relatively long time baseline help us to constrain the orbital periods accurately.
With orbital periods of $1.0-1.1$ hr, J1053+5200 and J1436+5010 are the shortest period binary WDs yet discovered
\citep[see also][]{mullally09}.

\begin{figure}
\plottwo{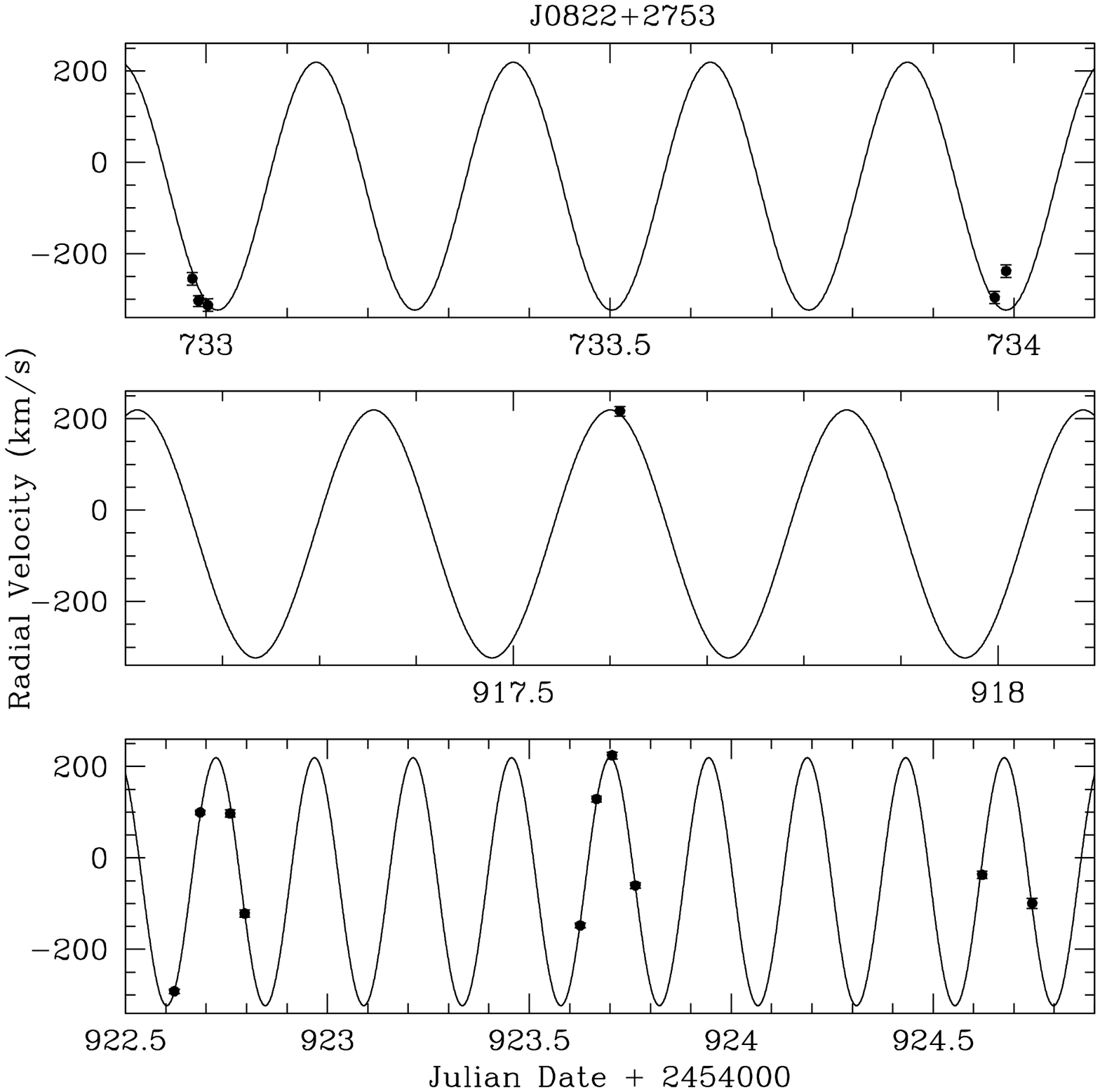}{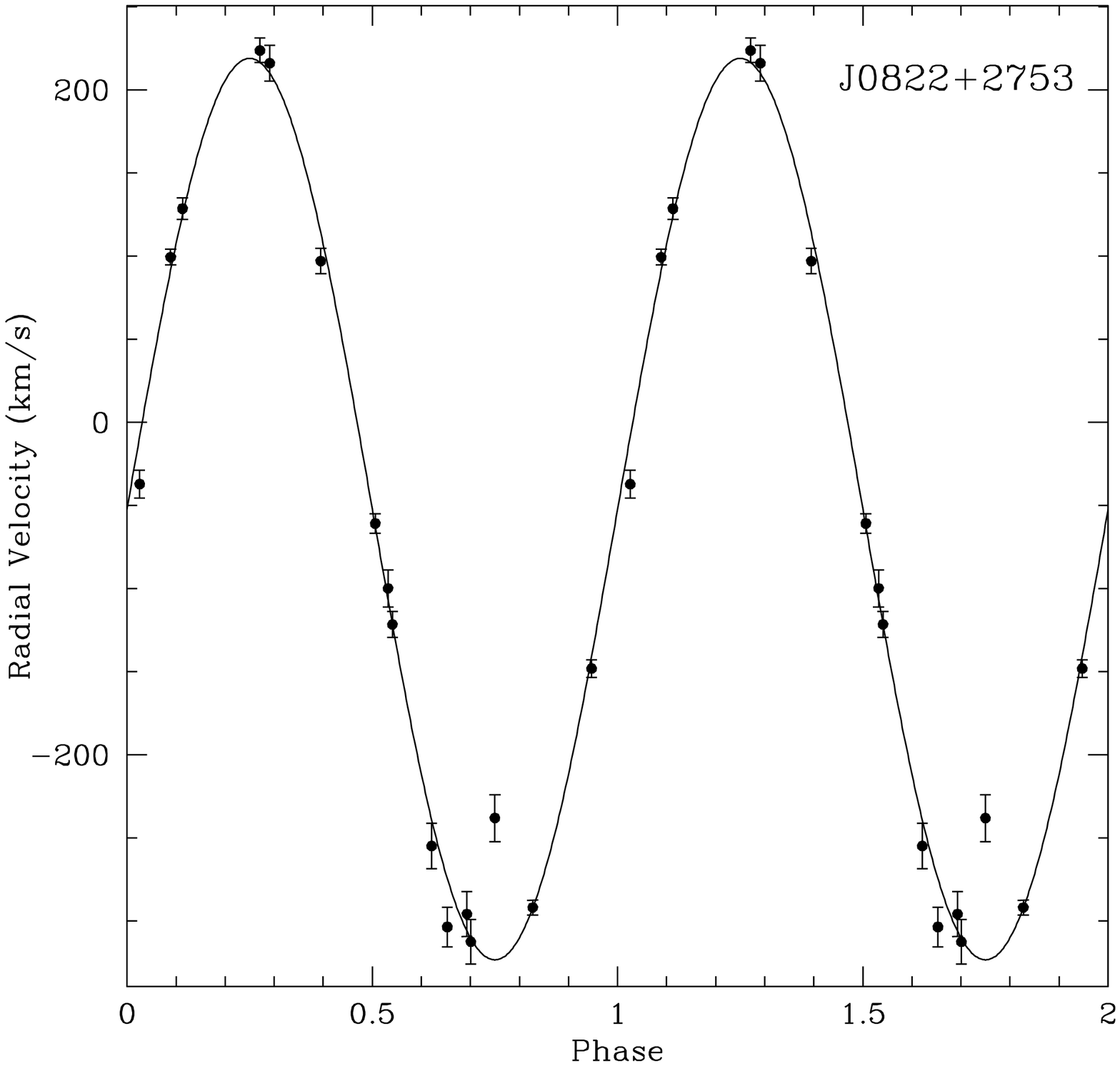}
\caption{The radial velocities of the white dwarf J0822+2753 observed in 2008 September (top left panel), 2009 March (middle left panel)
and 2009 April (bottom left panel). The right panel shows all of these data points phased with the best-fit period. 
The solid line represents the best-fit model for a circular orbit with a radial
velocity amplitude of 271.1 km s$^{-1}$ and a period of 0.2440 days.}
\end{figure}

\begin{figure*}
\plottwo{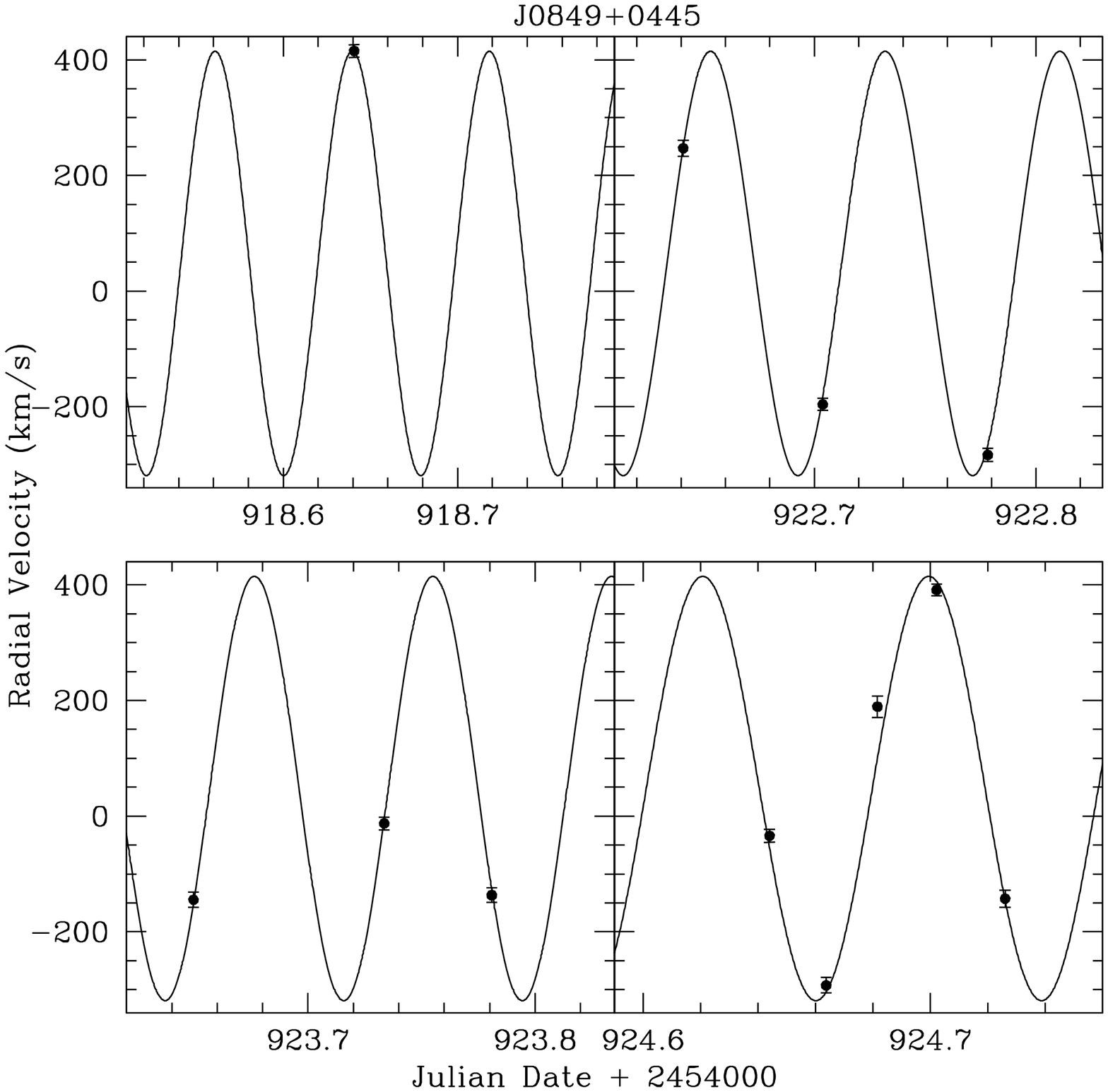}{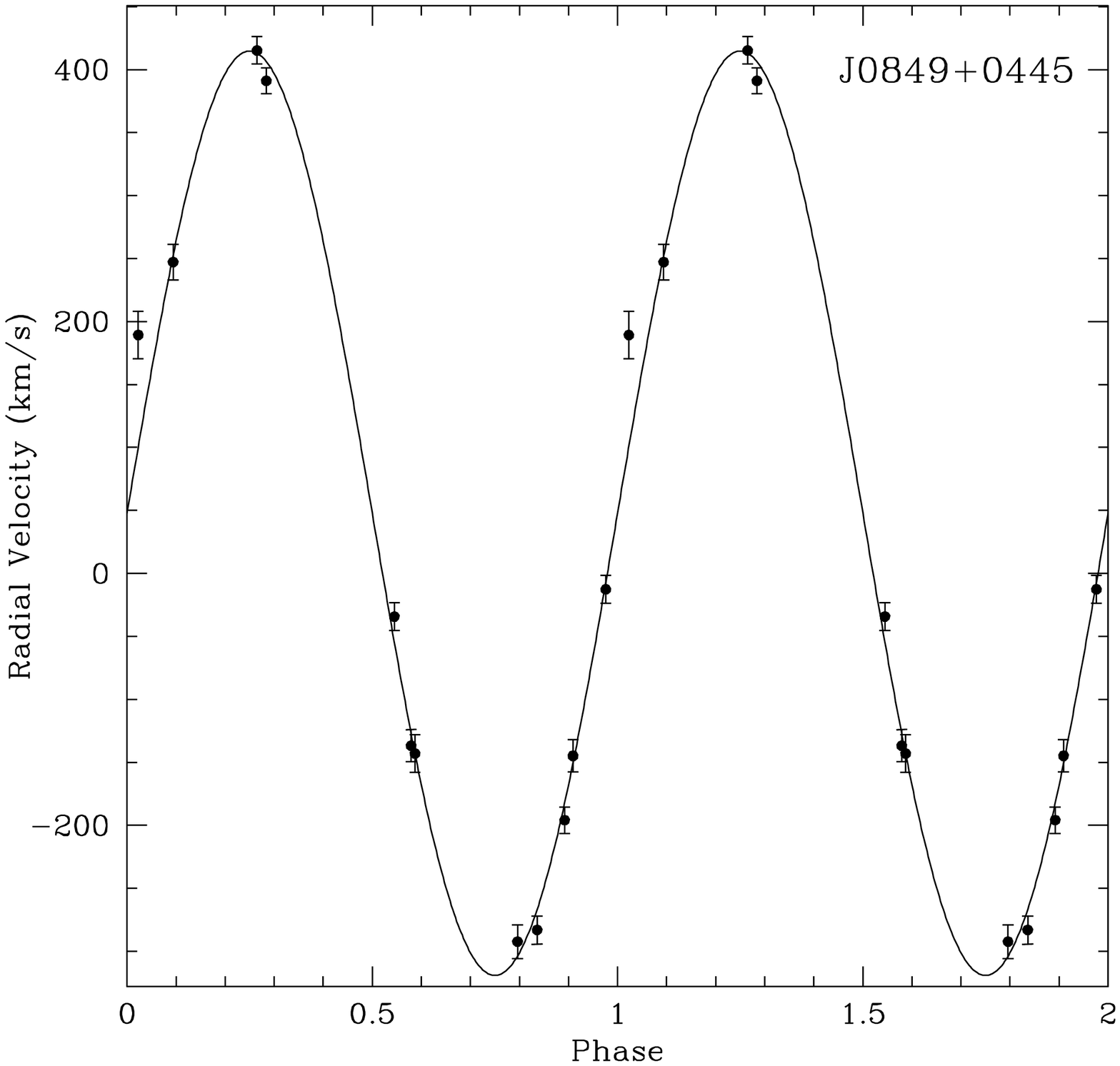}
\caption{The radial velocities of the white dwarf J0849+0445 observed in 2009 March and April (left panels). 
The right panel shows all of these data points phased with the best-fit period.
The solid line represents the best-fit
model for a circular orbit with $K=366.9$ km s$^{-1}$ and $P=0.0787$ days.}
\end{figure*}

\begin{figure*}
\plottwo{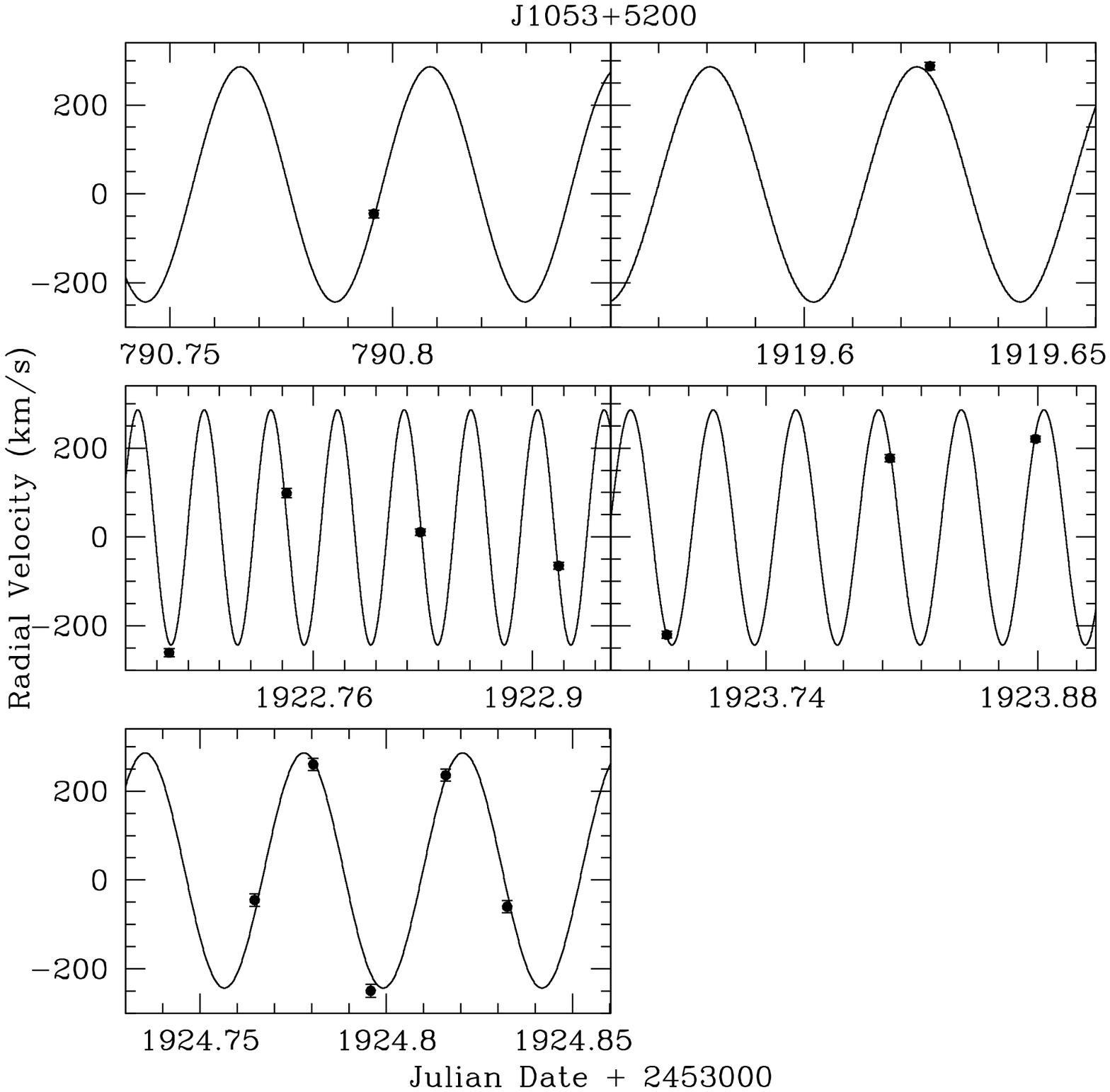}{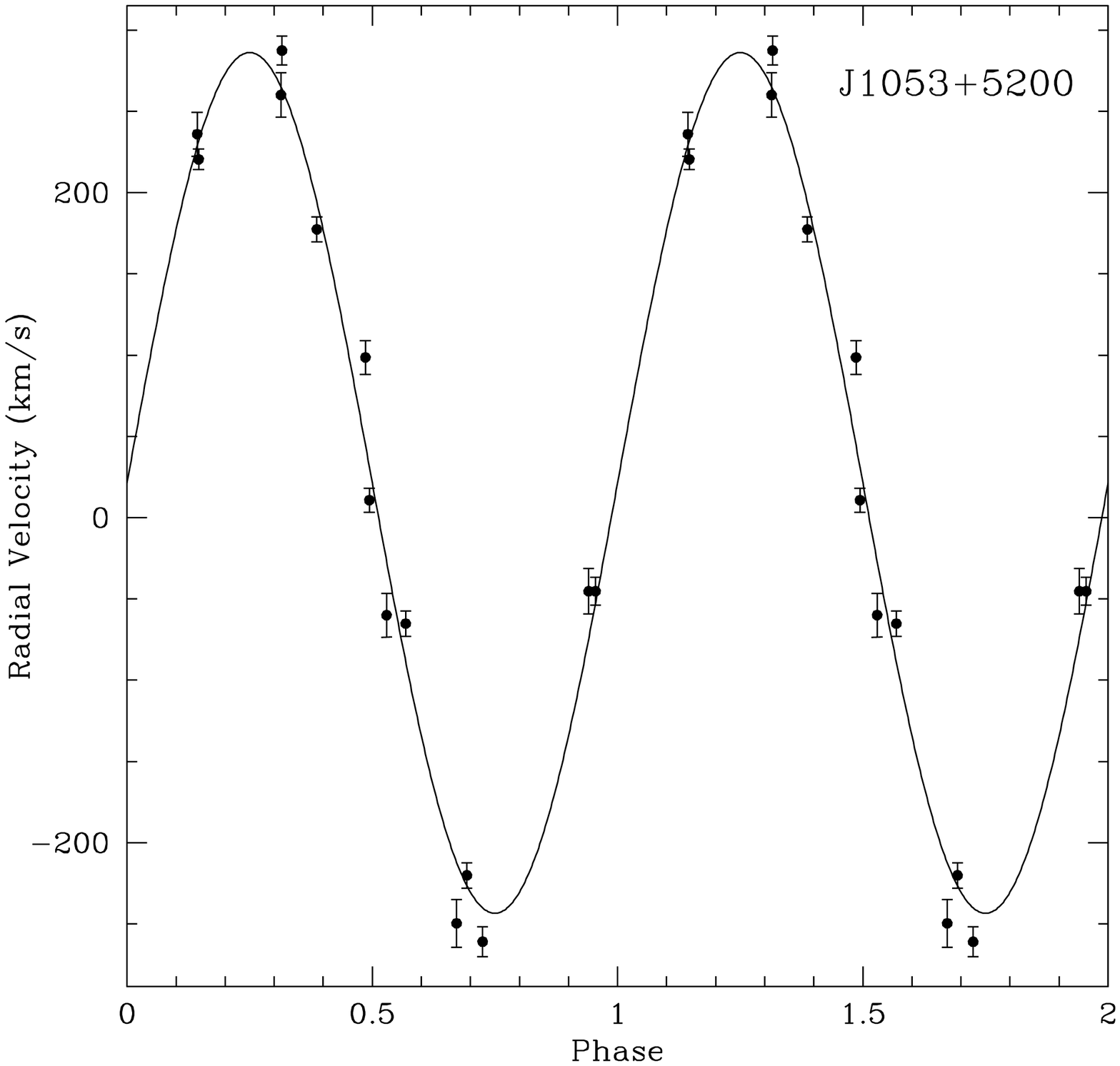}
\caption{The radial velocities of the white dwarf J1053+5200 observed in 2006 February, 2009 March and April (left panels). 
The right panel shows all of these data points phased with the best-fit period.
The solid line represents the best-fit model for a circular orbit with 
$K=264.8$ km s$^{-1}$ and $P=0.0426$ days.}
\end{figure*}

\begin{figure*}
\plottwo{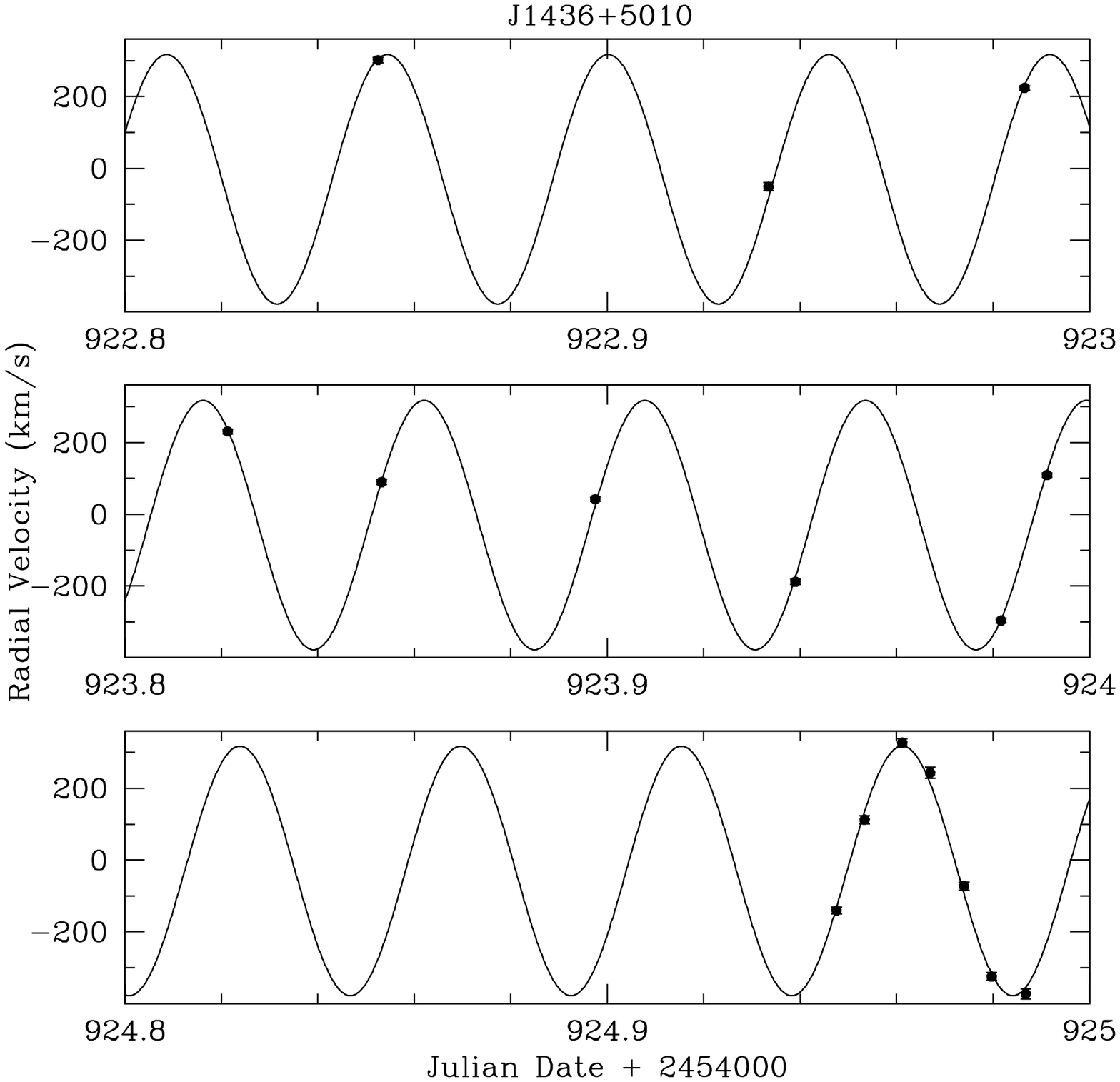}{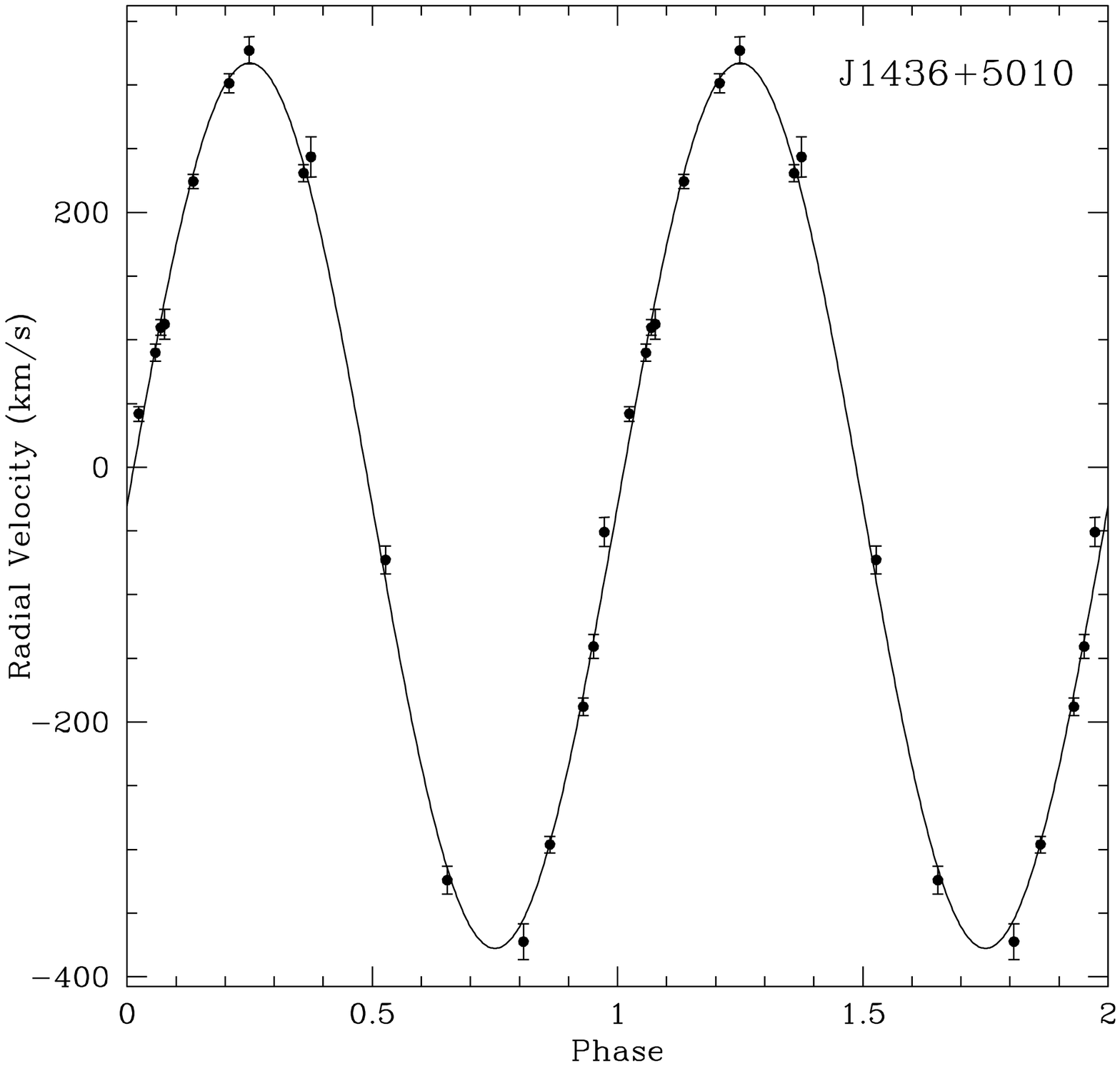}
\caption{The radial velocities of the white dwarf J1436+5010 observed in 2009 April (left panels).
The right panel shows all of these data points phased with the best-fit period.
The solid line represents the
best-fit model for a circular orbit with $K=347.4$ km s$^{-1}$ and $P=0.0458$ days.}
\end{figure*}

The orbital parameters for our targets, including the orbital period, semi-amplitude ($K$) of the radial velocity variations,
systemic velocity, the time of spectroscopic conjunction, and mass function, are presented in Table 2.
The systemic velocities in this table include
a gravitational redshift term, which should be subtracted from these velocities to find the true systemic velocities.
This correction is on the order of 3 km s$^{-1}$ for our targets (see the discussion below).

%TABLE2
\begin{deluxetable}{cclrcr}
\tabletypesize{\footnotesize}
\tablecolumns{6}
\tablewidth{0pt}
\tablecaption{Orbital Parameters}
\tablehead{
\colhead{Object}& 
\colhead{$P$}& 
\colhead{$K$}&
\colhead{Systemic Velocity}&
\colhead{Spectroscopic Conjunction}& 
\colhead{Mass Function}\\
  & (days) & (km s$^{-1}$) & (km s$^{-1}$) & (HJD)  &
}
\startdata 
J0822+2753 & 0.2440 $\pm$ 0.0002 & 271.1 $\pm$ 9.0  & $-$52.2 $\pm$ 4.5 & 2454732.8312 & 0.5038 $\pm$ 0.07745 \\
J0849+0445 & 0.0787 $\pm$ 0.0001 & 366.9 $\pm$ 14.7 &   47.8  $\pm$ 7.4 & 2454918.6200 & 0.4026 $\pm$ 0.06380 \\
J1053+5200 & 0.0426 $\pm$ 0.0001 & 264.8 $\pm$ 15.0 &   21.4  $\pm$ 7.7 & 2453790.7977 & 0.08195 $\pm$ 0.01418 \\
J1436+5010 & 0.0458 $\pm$ 0.0001 & 347.4 $\pm$ 8.9  & $-$30.2 $\pm$ 5.1 & 2454922.8430 & 0.1990  $\pm$ 0.02969 \\
\enddata 
\end{deluxetable}

We perform model fits to each individual spectrum and also to the average composite spectra using synthetic DA WD spectra kindly
provided by D. Koester. We use the individual spectra to obtain a robust estimate of the errors in our analysis.
Figure 5 shows the composite spectra and our fits using the entire wavelength range. The best-fit $T_{\rm eff}$ and
$\log g$ values are given in Table 3.
We obtain best-fit solutions of $8880-16550$ K and $\log g= 6.23-6.69$ for our targets,
confirming that they are ELM WDs.

\begin{figure*}
\plottwo{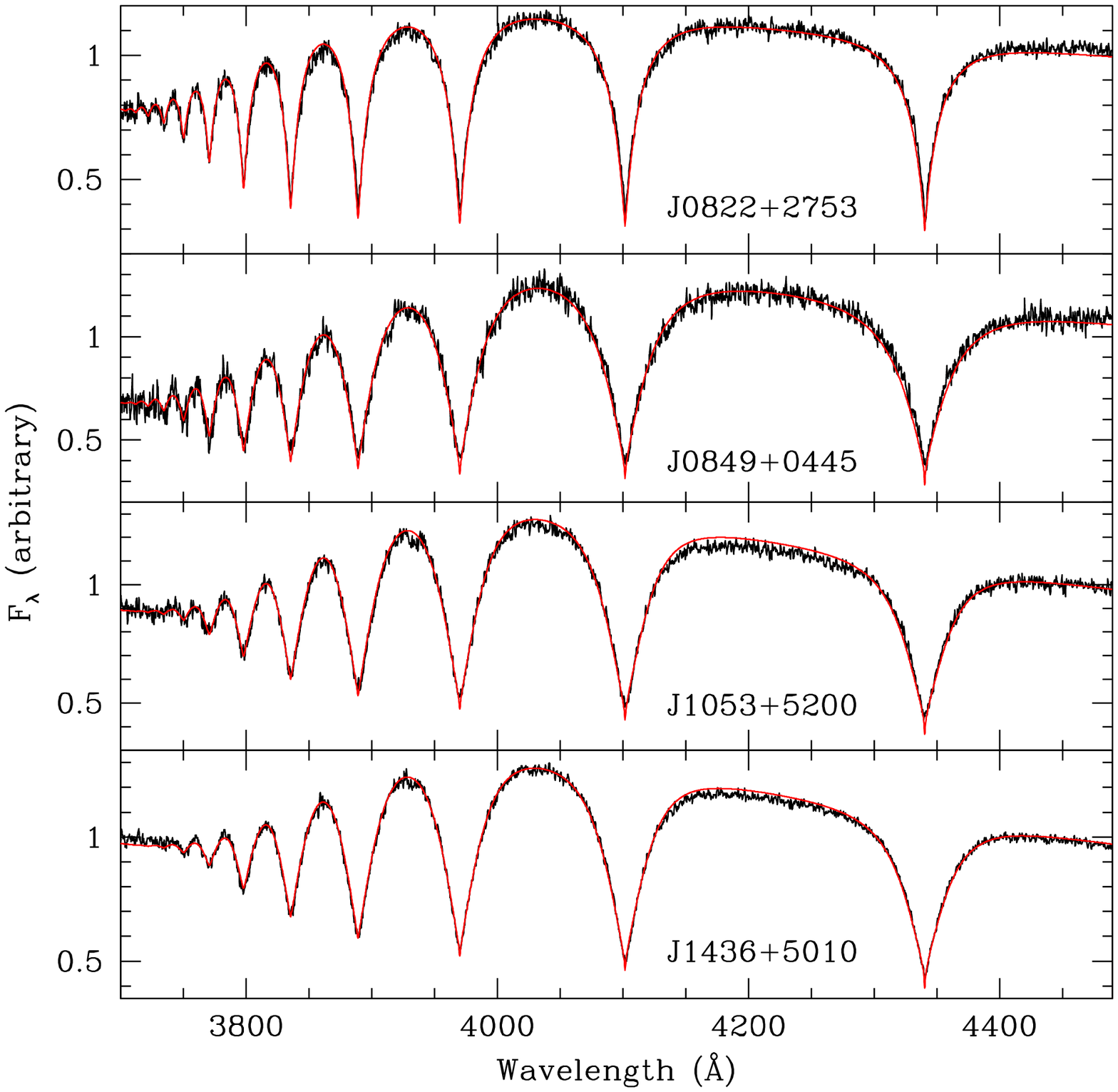}{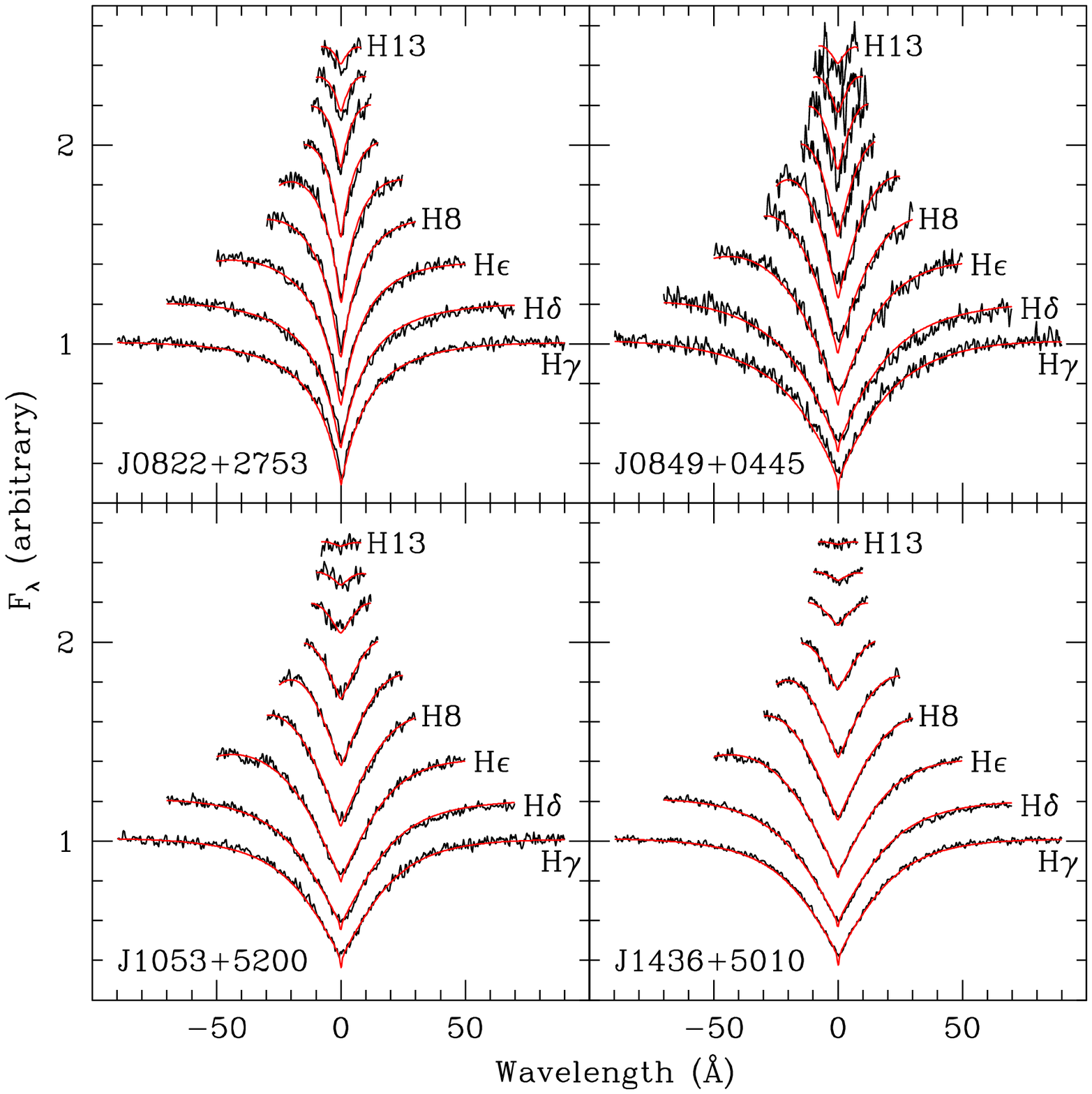}
\caption{Spectral fits (red solid lines) to the composite spectra of our targets (jagged lines, left panels)
and to the flux-normalized line profiles (right panels).}
\end{figure*}

%TABLE3
\begin{deluxetable}{ccccccccc}
\tabletypesize{\scriptsize}
\tablecolumns{9}
\tablewidth{0pt}
\tablecaption{Physical Parameters}
\tablehead{
\colhead{Object}&
\colhead{$T_{\rm eff}$}&
\colhead{$\log g$}&
\colhead{Mass}&
\colhead{Companion Mass}&
\colhead{$i=60^\circ$}&
\colhead{NS}&
\colhead{SN Ia}&
\colhead{Merger Time}\\
  & (K) &  & ($M_\odot$) & ($M_\odot$) & ($M_\odot$) & Probability & Probability & (Gyr)
}
\startdata
J0822+2753 & 8880  $\pm$ 60  & 6.44 $\pm$ 0.11 & 0.17 & $\geq0.76$ & 1.05 & 18\% & 5\% & $\leq8.42$ \\
J0849+0445 & 10290 $\pm$ 250 & 6.23 $\pm$ 0.08 & 0.17 & $\geq0.64$ & 0.88 & 15\% & 4\% & $\leq0.47$ \\
J1053+5200 & 15180 $\pm$ 600 & 6.55 $\pm$ 0.09 & 0.20 & $\geq0.26$ & 0.33 &  4\% & 1\% & $\leq0.16$\\
J1436+5010 & 16550 $\pm$ 260 & 6.69 $\pm$ 0.07 & 0.24 & $\geq0.46$ & 0.60 &  9\% & 4\% & $\leq0.10$\\
\enddata
\end{deluxetable}

Slight differences between the continuum level of the observations and that of the best-fit model spectrum
are evident for J1053+5200. These differences are suggestive of an imperfect flux calibration.
If we normalize (continuum-correct) the
composite spectra and fit just the Balmer lines, we obtain best-fit solutions that differ by $70-800$ K in $T_{\rm eff}$ and
0.1 dex in $\log g$ for our four targets. These fits are shown in the right panel of Figure 5.
These solutions are consistent with the fits to the entire spectra within the errors.
Our results are consistent with the \citet{eisenstein06} analysis within 500 K in temperature and 0.4 dex in surface gravity.
Similarly, they are also consistent with the \citet{mullally09} analysis within 1000 K in temperature and 0.2 dex in surface
gravity.
Given the higher resolution and higher signal-to-noise ratio MMT data, shorter exposure times, and extended blue coverage
that includes gravity-sensitive higher order Balmer lines, our $T_{\rm eff}$,
$\log g$, and orbital period estimates (Table 2 and 3) should be more reliable.

Figure 6 shows the effective temperatures and surface gravities for our targets (red circles) plus the previously
identified ELM WDs in the literature. Open circles show the WD companions to milli-second pulsars PSR J1012+5307 and J1911-5958A
\citep{vankerkwijk96,bassa06}. Filled triangles show the ELM sdB star HD 188112 \citep{heber03}, and the WDs SDSS J0917+4638
\citep{kilic07a,kilic07b} and LP400--22 \citep{kawka06,kilic09,vennes09}. All of these WDs show radial velocity variations.
Solid lines show the constant mass tracks for low mass WDs from our updated model calculations based on
the \citet[][labeled in $M_\odot$ on the right side of the figure]{panei07} study.
We model the evolution of He-core WDs in close binary systems, improving on the earlier models
of \citet{althaus01} and \citet{panei07}.
Instead of $M\geq0.18M_\odot$ as found in \citet{althaus01}, we find that for masses
$M>0.17017~M_\odot$ diffusion-induced hydrogen-shell flashes take place, which yield small hydrogen envelopes.
The models with $M \leq 0.17017~M_\odot$ do not experience
thermonuclear flashes. As a result, they have massive hydrogen envelopes, larger radii,
lower surface gravities, and they evolve much more slowly compared to more massive WDs.

\begin{figure}
\plotone{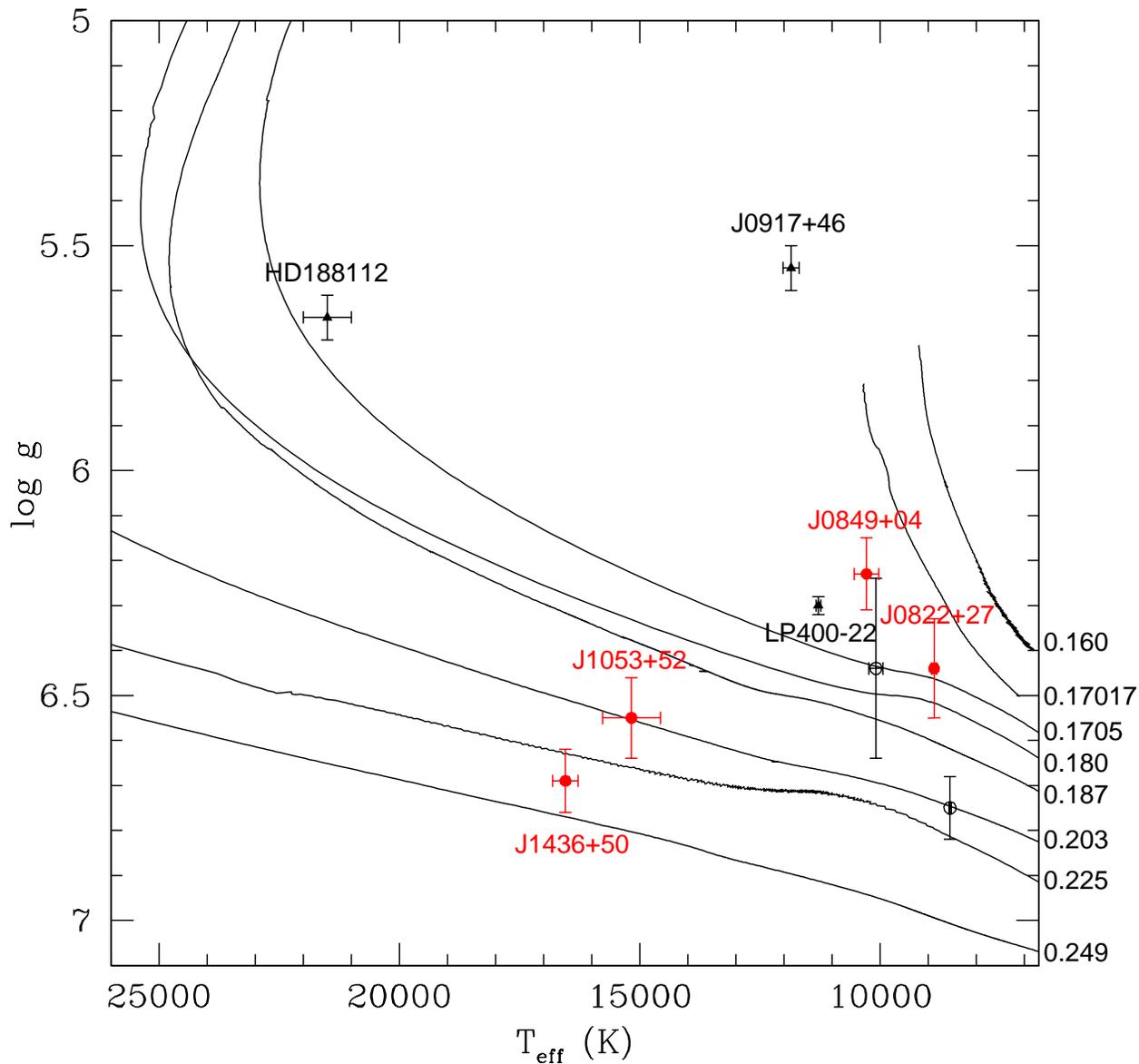}
\caption{The best fit solutions for the surface gravity and temperatures of our targets (filled circles),
overlaid on tracks of constant mass from our new calculations based on the \citet{panei07} models.
Spectroscopically confirmed ELM WDs and sdBs found in the literature are shown as
filled triangles. Companions to milli-second pulsars PSR J1012+5307 and J1911-5958A are shown as open circles.}
\end{figure}

Figure 6 shows that both J0822+2753 and J0849+0445 are $0.17~M_\odot$ WDs, whereas J1053+5200
and J1436+5010 are $0.20~M_\odot$ and $0.24~M_\odot$ WDs, respectively. The errors in these mass estimates are
$\approx 0.01~M_\odot$. The models by \citet{althaus01} also predict masses within 0.01-0.02 $M_\odot$ of the above estimates.

\section{DISCUSSION}

\subsection{J0822+2753}

The temperature and surface gravity for J0822+2753 imply an absolute magnitude of
$M_{\rm g}=10.1$ mag, a radius of $0.04~R_\odot$, and an age of 1.2 Gyr. This
luminosity places it at a distance of 430 pc, 250 pc above the plane.
J0822+2753 has a proper motion of $\mu_{\alpha} cos \delta = 3.4 \pm 3.5$ and $\mu_{\delta}=-19.2 \pm 3.5$ mas yr$^{-1}$
\citep{munn04}. Based on the mass and radius estimates, the gravitational redshift of the WD
is 2.7 km s$^{-1}$. Therefore, the true systemic velocity is $-54.9$ km $^{-1}$.
The velocity components with respect to the local standard of rest as defined by \citet{hogg05} are
$U = 64 \pm 6, V = -22 \pm 8$, and $W = -25 \pm 7$ km s$^{-1}$. Clearly, J0822+2753 is
a disk star \citep{chiba00}.

We combine the spectra near maximum blue-shifted radial velocity and near minimum radial velocity into two composite spectra. If there
is a contribution from a companion object, it may be visible as an asymmetry in the line profiles.
We do not see any obvious asymmetries in the line profiles and our optical spectroscopy does not reveal any spectral
features from a companion object. A main-sequence star companion with $M\geq0.76~M_\odot$
would have $M_I<6.5$ mag \citep{kroupa97}, brighter than the low-mass WD ($M_I\approx 9.3$ mag) and detectable in the $I$-band.

Figure 7 shows the SDSS photometry of our targets compared to WD model predictions.
None of the targets in our sample, including J0822+2753, shows excess flux in the optical.
Hence, a main-sequence star companion is ruled out for J0822+2753.

\begin{figure}
\plotone{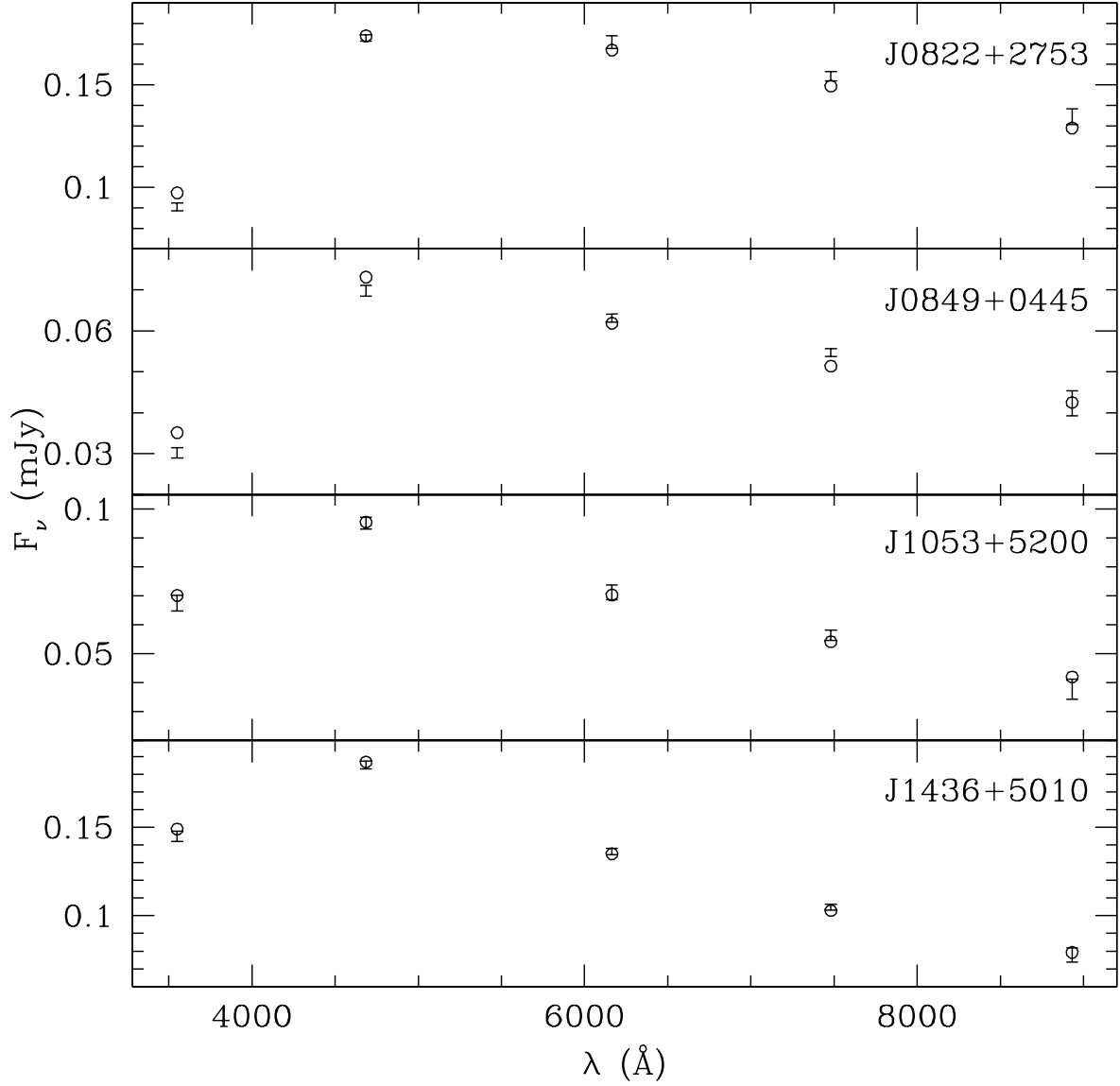}
\caption{Spectral energy distributions of our targets (error bars) and the WD model predictions (open circles).}
\end{figure}

Using the mean inclination angle for a random stellar sample, $i=60^{\circ}$, the companion has a mass of
$1.05~M_\odot$ and a separation of $1.8~R_\odot$. This separation is about $40\times$
larger than the radius of the WD. The probability of a neutron star companion, i.e. $M=1.4-3.0~M_\odot$, is 18\%.
\citet{agueros09a} obtained radio observations of all four of our targets, including J0822+2753, using the Green Bank Telescope. 
They do not see any evidence of a milli-second pulsar companion. Therefore, the companion is most likely a massive WD. However,
follow-up x-ray observations are required to rule out a neutron star companion.

\subsection{J0849+0445}

J0849+0445 falls in between the cooling sequences for $M=0.17017$ and 0.1705 $M_\odot$ in the temperature versus surface
gravity diagram (Figure 6).
Thus, J0849+0445 is 1-2 Gyr old, has $M_g\approx9.5$ mag,
and is located at 930 pc away from the Sun. The relatively large uncertainty in the age estimate is due to the differences
between the models with masses larger/smaller than 0.17017 $M_\odot$.
The measured proper motion for J0849+0445 is consistent with zero \citep{munn04}.
The gravitational redshift of the WD is $\approx$2.3 km s$^{-1}$. Hence, the true systemic velocity is 45.5 km s$^{-1}$.
The velocity components with respect to the local standard of rest are $U=-21 \pm 13, V=-24 \pm 13$, and $W=25 \pm 14$ km s$^{-1}$.
J0849+0445 is also a disk star.

As in the case of J0822+2753, we do not see any evidence of a companion in our spectra. Main-sequence companions are ruled out
based on the SDSS photometry. In addition, radio observations do not reveal a milli-second pulsar companion.
There is a 70\% probability
that the companion is less massive than $1.4~M_\odot$. Therefore, the companion is most likely another WD with $M\geq0.64~M_\odot$.
Assuming an inclination angle of $60^{\circ}$, the companion is probably a $0.88~M_\odot$ object
at $0.8~R_\odot$ (or $15 \times R_{WD}$) separation. 

\subsection{J1053+5200}

J1053+5200 has $M=0.20~M_\odot, R=0.04~R_\odot$, and $M_{\rm g}=8.7$ mag. This absolute magnitude corresponds to a WD cooling
age of 160 Myr. Its distance is 1.1 kpc, 1 kpc above the plane.
It has a proper motion of $\mu_{\alpha} cos \delta = -29.7 \pm 3.5$ and $\mu_{\delta}=-31.2 \pm 3.5$ mas yr$^{-1}$
\citep{munn04}. The gravitational redshift of the WD is 3.2 km s$^{-1}$. The velocity components with respect to the
local standard of rest are $U=-110 \pm 19, V=-192 \pm 28,$ and $W=8 \pm 12$ km s$^{-1}$. 
J1053+5200 lags by $\sim 200$ km s$^{-1}$ behind the Galactic disk and is clearly a halo star.

The mass function for J1053+5200 implies that the companion is more massive than $0.26~M_\odot$. A $0.26~M_\odot$ main-sequence
star would be about 50\% fainter than the WD in the $I-$band and it would have been detected in the SDSS $i-$ and $z-$ band data.
The lack of excess flux in the SDSS photometry (Figure 7) rules out a main-sequence companion.
A neutron star companion is unlikely (4\% probability).
The companion is most likely another WD, and specifically a low-mass ($M<0.45~M_\odot$) WD.
J1053+5200 has a 70\% chance of having a low-mass WD companion.
Assuming an inclination angle of $60^{\circ}$, the companion is probably a
low mass object with $M = 0.33~M_\odot$ at $0.4~R_\odot$ separation. 

\subsection{J1436+5010}

J1436+5010 is a $40-70$ Myr old $M\approx0.24 M_\odot$ WD at a distance of $\approx800$ pc, 710 pc above
the Galactic plane. The gravitational
redshift of the WD is 4.2 km s$^{-1}$. It has a proper motion of $\mu_{\alpha} cos \delta = 7.8 \pm 3.5$ and 
$\mu_{\delta}=-5.1 \pm 3.5$ mas yr$^{-1}$ \citep{munn04}. The velocity components with respect to the local standard of rest are
$U=44 \pm 14, V=-5 \pm 12$, and $W=-28 \pm 8$ km s$^{-1}$. J1436+5010 is a disk object.

We do not see any evidence of a companion in our MMT spectra of this object. In addition, the companion has to be more massive than
$0.46~M_\odot$, and such a main-sequence star companion is ruled out based on the SDSS photometry data. There is a 9\% chance that
the companion is a neutron star. However, the companion is most likely a carbon/oxygen core WD.
Assuming an inclination angle of $60^{\circ}$, the companion is probably a
$0.60~M_\odot$ object at $0.5~R_\odot$ separation. 

\citet{ramirez07} used Gaussian probability distributions to
assign stars to the different Galactic components based on
kinematics. Using the same criterion, J0822+2753, J0849+0445, and J1436+5010 have 94-97\% chances of
being thin disk members. On the other hand, J1053+5200 has a 99\% chance of being a halo member.
These statistics confirm our membership assignments presented above.

\section{THE COMMON-ENVELOPE PHASE}

Our radial velocity observations and the available optical photometry show that none of our targets have main-sequence
companions. The probability of neutron star companions ($M>1.4~M_\odot$) ranges from 4\% to 18\%. 
However, no such companions are visible in the radio data \citep{agueros09a}.
Therefore, the companions are most likely other WDs.

The formation scenarios for close WD pairs include two consecutive common-envelope phases or an Algol-like
stable mass transfer phase followed by a common-envelope phase \citep[e.g.][]{iben97}.
The mass ratios ($q=M_{\rm bright}/M_{\rm dim}$) for our targets range from 0.22 to 0.77, and they favor
a scenario involving two common-envelope phases \citep[see][and references therein]{nelemans00}. 
 
\citet{nelemans00} and \citet{net05} demonstrate that 
the standard common-envelope formalism (the $\alpha$-formalism equating the energy balance in the system) does not always work.
Instead, they suggest that the common-envelope evolution of close WD binaries can be reconstructed with an
algorithm ($\gamma$-algorithm) imposing angular momentum balance. 
Studying the prior evolution
of 10 WD+WD binaries where both WD masses are known, \citet{nelemans05} find that the $\gamma$-algorithm
is able to explain the observed properties of those systems following two common-envelope phases.
Most of the systems can be accounted for with a single-valued $\gamma=1.5$, where $\gamma$ is the
rate of angular momentum loss as defined by \citet{paczynski67}. However, $\gamma$ is not a stiff parameter, and
it may vary between 0.5 and 4. The same systems may be explained by means of various $\gamma$s and assuming
different initial parameters \citep{net05}.
The mass ratios of these 10 systems are on the order of unity.

\citet{kilic07b,kilic09} used the $\gamma$-algorithm to study the prior evolution of the
ELM WD systems J0917+4638 and LP400--22. These two systems have mass ratios of $\leq0.61$ and $\leq0.46$, respectively.
Using $\gamma=1.5$, \citet{kilic07b} find that the J0917+4638 system can be explained as the descendant of a binary system including a
2.2 and a 0.8 $M_\odot$ star at an orbital separation of 0.4 AU. A similar study of the LP400-22 system shows that
$\gamma=2$ may be more appropriate for this binary \citep{kilic09}.

In order to understand the prior history of the short-period binary systems presented in this paper,
we consider an initial system involving a 3 $M_\odot$ and a 1 $M_\odot$ star.
The 3 $M_\odot$ star evolves off the main-sequence, overflows its Roche lobe as a giant with a 0.6 $M_\odot$ core,
forms a helium star (sdB) which does not expand after He-exhaustion in the core, and turns into a WD.
The 1 $M_\odot$ star also overflows its Roche lobe after main-sequence evolution, when its core is 0.24 $M_\odot$.
We also assume that the current orbital period is 0.0458 days (as in the J1436+5010 system). 
The current separation of this system is 0.5 $R_\odot$. Evolving the system back in time, and considering
the latest common-envelope phase, the 1 $M_\odot$ star had a 0.24 $M_\odot$ core when its radius
was 10 $R_\odot$ \citep{iben85}. Taking this radius as the Roche lobe radius \citep{eggleton83}, the separation prior
to the second common-envelope phase is 25 $R_\odot$ \citep{net05}, which corresponds to $\gamma=2.01$.

Considering the first common-envelope phase, the 3 $M_\odot$ star has a 0.6 $M_\odot$ core when its radius
is 410 $R_\odot$. Taking this radius as the Roche lobe radius, the separation prior to
the first common-envelope phase is 860 $R_\odot$ (4.0 AU), which corresponds to $\gamma=1.58$. 
Similar to this sytem, the common-envelope evolution of all four systems discussed in this paper can be explained
with $\gamma=1.6 - 2.2$.
In this scenario, the companions are older, smaller, and fainter than the $0.17-0.24 M_\odot$ WDs observed today, consistent with the
lack of evidence for the presence of companions in the SDSS photometry and our optical spectroscopy.

\section{THE FUTURE: MERGER PRODUCTS}

Short period binaries may merge within a Hubble time by losing angular momentum through gravitational radiation. 
The merger time for such binaries is

\begin{equation}
\tau \approx \frac{(M_1 + M_2)^{1/3}}{M_1 M_2} P^{8/3} \times 10^7 yr
\end{equation}

\noindent where the masses are in solar units and the period is in hours \citep{landau58}.
For minimum mass companions ($i=90^\circ$) the merger times for our targets (in right ascension order)
are 8420 , 470, 160, and 100 Myr, respectively. All four targets will merge within a Hubble time.
We now explore possible outcomes of the merger process.

\subsection{EXTREME HELIUM STARS}

Hydrogen-deficient luminous stars, in order of decreasing effective temperature, include extreme helium, RCrB,
and hydrogen-deficient carbon stars. Studies of the chemical compositions of these stars suggest that they
form an evolutionary sequence \citep{hernandez09}. There are two leading scenarios to explain the origin
of extreme helium stars. In one scenario, the merger of a He-star or a He-core WD with a carbon/oxygen core WD
forms a hydrogen-deficient supergiant \citep{webbink84,iben84,iben96,saio02}. The other scenario, commonly referred
to as the born-again scenario, suggests that
a hydrogen-deficient star forms when a post-asymptotic giant branch star experiences a late helium shell flash
\citep[see][and references therein]{iben83}. This flash converts the hydrogen-rich envelope to helium, creating a hydrogen-deficient star.

Studies of elemental and isotopic abundances for carbon, nitrogen, and oxygen are useful for differentiating
between these two scenarios. 
Based on the observed abundances, \citet{saio02} and \citet{pandey06} argue that most hydrogen-deficient carbon
stars and RCrB stars form through WD mergers.
In addition, \citet{clayton07} find that every RCrB and hydrogen-deficient carbon star that they
have observed has enhanced $^{18}$O/$^{16}$O ratios compared to the solar value. They propose the WD merger
scenario as a likely formation mechanism. Based on preliminary calculations, \citet{clayton07}
suggest that the accretion of the helium WD by the carbon/oxygen WD is rapid and it induces nucleosynthesis, which
converts $^{14}$N to $^{18}$O by $\alpha$-capture. 

The overproduction of
$^{18}$O is not predicted in the born-again scenario, because either the $^{14}$N is burnt
to $^{22}$Ne or $^{18}$O is destroyed by proton capture \citep{hernandez09}.
There is at least one star\footnote{Two more have been discussed extensively in the literature,
FG Sge and V605 Aql.} identified as a product of the born-again scenario \citep[see][and references therein]{clayton06,miller07},
Sakurai's object (V4334 Sgr). Therefore, not all extreme helium stars form through WD mergers.

Three of the ELM WDs in our sample, J0822+2753, J0849+0445, and J1436+5010, have companions more massive
than 0.46 $M_\odot$. These 3 systems most likely have carbon/oxygen WD companions. They will merge within 100 Myr to
8.4 Gyr and form extreme helium stars\footnote{With a mass ratio of $\leq0.22$, J0822+2753 may become an AM CVn system
instead (see Section 6.5).}.
Therefore, these 3 systems provide independent evidence that
there is a mechanism to form hydrogen deficient stars through WD mergers.

\subsection{SINGLE HOT SUBDWARF STARS}

Close binary evolution plays an important role in the formation of subdwarf B stars, as witnessed by
the large fraction ($\geq40$\%) of sdB stars in binaries. Mass transfer between the companions and common-envelope
evolution can lead to large amount of mass loss prior to the start of core He-burning, creating an sdB star in a binary.
\citet{castellani93} demonstrated that if low-mass stars lose enough mass on the red giant branch, they can depart
the red giant branch before the core He flash. Instead, the He flash happens on the hot WD cooling track.
Depending on when this flash happens, sdB or He-enriched sdO stars form.

Only about 4\% of He-enriched sdO stars are in binary systems \citep{napiwotzki04b}.
The mergers of two He-core WDs can explain the lower fraction of binaries observed among
the He-enriched sdO stars \citep{han03}. With shrinking separation,
the less massive object is accreted onto the companion, leading to He ignition \citep{heber09}.
The merger product will be enriched in CNO, similar to the observed abundances of 
He-enriched sdO stars \citep{saio00}.
 
In addition to the merger scenario, single hot subdwarf stars can also be produced through a common-envelope
phase with a massive planet or a brown dwarf. The recent discoveries of planets around the subdwarf B
stars V Pegasi \citep{silvotti07} and HD 149382 \citep{geier09} show that
this channel of formation contributes to the single subdwarf B star population.

There is a 70\% chance that the companion to one of our targets, J1053+5200, is a low-mass He-core WD. The merger
of this binary WD system will most likely create a He-enriched sdO star. Hence, binary mergers of two He-core WDs
contribute to the single hot subdwarf population in the Galaxy.

\subsection{SNe Ia and .Ia}

SNe Ia are caused by the thermonuclear explosion of WDs growing to the Chandrasekhar mass either by accretion from
a companion or by mergers of two degenerate stars. The double degenerate scenario requires mainly
mergers of two CO WDs to have sufficient mass to exceed the Chandrasekhar mass limit \citep{webbink84,iben84}. 
However, if the mass transfer is stable, accretion from a He-core WD can also result in accumulation of Chandrasekhar
mass by the CO WD accretor and this can result in a SN Ia explosion.
The population synthesis models by
\citet{yungelson05} demonstrate that the expected contribution of He+CO WD binaries to the SNe Ia rate is two
orders of magnitude smaller than that of the CO+CO WD systems.

The visible components of the four binaries discussed in this paper are 0.17-0.24 $M_\odot$ WDs. There is
a 1-5\% chance that the companions are massive WDs and the total mass of the binary systems exceed 1.4 $M_\odot$
(see Table 3). Our targets are most likely not SNe Ia progenitors.

Recently, \citet{guillochon09} presented a new mechanism for the detonation of a sub-Chandrasekhar mass CO WD
through accretion from a low-mass He-core WD. If the mass accretion is dynamically unstable, the
instabilities in the accretion stream can lead to 
the detonation of surface helium that accumulates on the CO primary during the final few orbits prior to
merger. These detonations are likely to resemble dim Type Ia SNe \citep[or .Ia,][]{bildsten07}, and would primarily synthesize
intermediate-mass elements. Under certain conditions, the primary itself is also detonated.
An important feature of this mechanism is that the total system mass does not need to exceed the
Chandrasekhar limit. As such, the systems we present in this paper could very well be the progenitor
systems for these events.

\subsection{WHITE DWARF + NEUTRON STAR MERGERS}

Green Bank Telescope observations
do not detect milli-second pulsar companions around our targets.
However, neutron star companions cannot be ruled out based on the radio data alone.
X-ray observations can detect the blackbody emission from a neutron star companion even if it
is radio-quiet or if its pulsar beam misses our line of sight. XMM-Newton observations
of the extremely low mass WD, SDSS J0917+4638, do not detect a neutron star in the system
\citep{agueros09b}. Similar observations will be necessary to search
for neutron star companions in our targets.

\citet{nelemans01} estimate that NS + WD merger rate is two orders of magnitude
smaller than WD + WD merger rate. However,
based on the mass function alone, there is a 4-18\% chance that the companions to our targets
are neutron stars. 
\citet{garcia07} study the evolution of a 1.4 $M_\odot$ neutron star and a 
0.6 $M_\odot$ merger using a smooth particle hydrodynamics code.
They find that, once the white dwarf has filled its Roche lobe, it is disrupted in a few orbital periods,
e.g. on the order of 5 minutes.
The final configuration consists of a neutron star surrounded by an accretion disk, and the mass
loss from the system is negligible. 
However, the outcome of mergers of NS with lighter WDs is not clear.
\citet{yungelson02} suggest that mass exchange in a 0.2 + 1.4 $M_\odot$ system may be stable.
\citet{nelemans10} argue for WD or He-star donors in several ultra-compact x-ray binary systems. These donors
have initial masses
larger than 0.32-0.35 $M_\odot$ and they somehow survive disruption in the merger with the NS. Therefore,
0.2 $M_\odot$ WDs may also survive mergers with neutron stars.

\subsection{AN ALTERNATIVE FUTURE: AM CVn SYSTEMS}

AM CVn stars are interacting double stars with WD accretors
and orbital periods less than about one hour. There are
three formation scenarios for AM CVn systems involving three
types of donor stars; WDs, helium stars, or evolved main-sequence stars
\citep[see][and references therin]{postnov06}.
Studying the CNO and He abundances of known AM CVn systems, \citet{nelemans10} find
evidence of WD donors in some systems, and evolved helium star donors
in others.

The WD channel requires a binary system with short enough orbital period that
gravitational wave radiation drives the stars into contact. The low-mass WD
fills its Roche lobe and transfers mass to the companion.
Depending on the mass ratio of the binary system (if the mass ratio is small),
the mass transfer is stable and the system evolves to longer periods
\citep[see][and references therein]{marsh04,nelemans10}. \citet{marsh04} argue that
despite the absence of a single system of extreme mass ratio amongst the observed close
double WD population, this channel of formation is probable.

Here, we have uncovered four binary systems with extreme mass ratios of $\leq0.22$ to
$\leq0.77$. The important question is whether
these systems will merge or if they will instead
create AM CVn systems. \citet{marsh04} suggest that the mass transfer between double
WDs can be dynamically stable, unstable, or the intermediate case of either stability
or instability depending on the degree of spin-orbit coupling.
\citet{motl07} and \citet{racine07} demonstrate that
the spin/orbit coupling is strong, raising the critical mass ratio to avoid merger
from around 0.2 (if there is no coupling) to 0.4-0.7. Based on these studies, we would
expect the majority of the WDs discussed in this paper to form AM CVn systems. However,
the SPH and grid-based calculations of mergers of WDs do not completely agree
on the outcome of contact, and the prior evolution of AM CVn systems is still uncertain
\citep[see][for a discussion]{fryer08}.
Nevertheless, with mass ratios of $\leq0.27$, J0822+2753 and J0849+0445 are strong candidates
for future AM CVn systems.

\section{CONCLUSIONS}

Almost all known double WD systems have mass ratios on the order of unity \citep{nelemans05}.
Recently, \citet{kilic07b,kilic09}
identified two WD binary systems with mass ratios $\leq0.6$. However, these two
systems (SDSS J0917 and LP400$-$22) will not merge within a Hubble time.

The four binary systems discussed in this paper will merge within a Hubble time.
They have extreme mass ratios and are likely progenitors of RCrB stars,
single He-enriched sdO stars, or AM CVn stars. These systems may even contribute
to the SNe Ia population, but the probability of this event is small. 

\citet{liebert05} estimate that low-mass ($M<0.45 M_\odot$) WDs make up about 10\%
of the local WD population, corresponding to a formation rate of 4 $\times 10^{-14}$ pc$^{-3}$ yr$^{-1}$.
\citet{eisenstein06} discovered only 13 ELM WDs out of a sample of 9316 WDs found in a survey volume
of $\geq$ 4 kpc$^3$ in the
SDSS Data Release 4 footprint. ELM WDs are rare, they make up about 0.14\% of the local WD sample.
Of course, the SDSS spectroscopic sample suffers from selection bias and incompleteness, but taken at face value,
0.14\% corresponds to a formation rate of 6 $\times 10^{-16}$ pc$^{-3}$ yr$^{-1}$.
However, all four ELM WDs discussed in this paper are going to merge within a Hubble time, and some within
the next few hundred Myr. Therefore, the formation rate of ELM WDs may be higher by an order of magnitude.
For the SDSS DR4 survey volume of 4 kpc$^3$, the formation rate is $0.2-2
\times 10^{-5}$ yr$^{-1}$, with the caveat that this number is highly uncertain due to selection bias present in the SDSS
spectroscopic survey.
\citet{bogomazov09} estimate a CO+He WD merger rate of 6-8 $\times 10^{-3}$ yr$^{-1}$.
The ELM WD mergers discussed
in this paper contribute only a small fraction to the overall CO+He WD merger rate in the Galaxy.
\citet{yungelson05} estimate a SNe Ia occurence rate of 10$^{-5}$ yr$^{-1}$ from CO+He WD progenitors. Our estimate
of the formation rate of ELM WDs is similar to this result.

In this paper, we present radial velocity data for four stars. However, we are obtaining
radial velocity observations for the remaining ELM WDs in the \citet{eisenstein06} sample.
To date, almost all of these ELM WDs show significant
radial velocity variations indicating the presence of a companion star. 
We are continuing to follow-up these objects at the MMT
to constrain their orbital periods accurately.
The importance of these observations is that, we not only find short period binaries that will merge within
a Hubble time, but also the majority of the binaries have extreme mass ratios. Understanding the prior history
of these systems requires an understanding of the common envelope phase. Studying the mass ratios and physical
characteristics of these systems will help in understanding common envelope evolution of close binary pairs.

\acknowledgements
Support for this work was provided by NASA through the {\em Spitzer Space Telescope} Fellowship Program,
under an award from Caltech. We thank M. Ag{\"u}eros, D. Steeghs, T. Marsh, and J. Guillochon for help discussions,
D. Koester for kindly providing DA WD model spectra, and an anonymous referee for a detailed and constructive report.

{\it Facilities:} \facility{MMT (Blue Channel Spectrograph)}

\end{document}